\documentclass[seceq]{ptptex}

\usepackage{graphicx}
\usepackage{wasysym}

\newcommand{\be}{\begin{equation}}
\newcommand{\ee}{\end{equation}}
\newcommand{\bea}{\begin{eqnarray}}
\newcommand{\eea}{\end{eqnarray}} 
\newcommand{\ld}{\ell_{\hbox{\tiny D}}}
\newcommand{\td}{\tau_{\hbox{\tiny D}}}

\title{
Random Deposition Model with a Constant Capture Length
}

\author{
Paolo \textsc{Politi$^{1,2,3,}$\footnote{E-mail: politi@ifac.cnr.it}} 
and Yukio \textsc{Saito$^{3}$}%
}

\inst{
$^1$ Istituto dei Sistemi Complessi, CNR, \\
Via Madonna del Piano, 50019 Sesto Fiorentino, Italy

$^2$ Istituto Nazionale per la Fisica della Materia, UdR Firenze, \\ 
Via G. Sansone 1, 50019 Sesto Fiorentino, Italy 

$^3$ Department of Physics, Keio University, 3-14-1 Hiyoshi,
Kohoku-ku, \\
Yokohama, 223-8522, Japan
}

\abst{
We introduce a sequential model for the deposition and aggregation of
particles in the submonolayer regime. 
Once a particle has been randomly deposited on the substrate, 
it sticks to the closest atom or island within
a distance $\ell$, otherwise it sticks to the deposition site.
We study this model both numerically and analytically in one dimension.
A clear comprehension of its statistical properties is provided,
thanks to capture equations and to the analysis of the
island-island distance distribution. 
}

\begin{document}

\maketitle

\section{Introduction}

The morphology of a system growing by deposition and aggregation
of particles~\cite{Meakin} depends on several factors
and it is not exaggerated to say that a detailed description
would require a theory, or at least a model, for each different physical
system. In spite of this, the use of simple and general models for
studying growth processes is widespread. This is justified by two big reasons:
First, such models allow to study large-scale features which are common
to different physical systems; Second, they allow to connect
apparently different phenomena.

In the following we are referring to the growth process of a crystal surface
by molecular beam epitaxy~\cite{MK}, with a special interest 
in the submonolayer regime~\cite{BE1992}, 
where only a fraction of a monolayer has been deposited. 
With these caveats in mind, a simple and very popular model for the
growth process~\cite{MK} accounts for random deposition of atoms 
and their thermally activated diffusion till they meet irreversibly
another atom (nucleation) or island (aggregation). The system is
therefore made up of diffusing particles (adatoms) and still islands.
New atoms, which are continuously provided by deposition, can not
leave the surface, because evaporation is forbidden at low temperature.
Hereafter, this model will be called `full diffusion' model.

The diffusion length, $\ld$, is an important quantity.
It measures the typical (linear) distance walked by an
adatom before being incorporated~\cite{PV}. 
It depends on the deposition rate, $F$,
and the hopping rate, $h$, through the relation
$\ld \sim (h/F)^{1/\gamma}$, with $\gamma=2(d^*+1)$, where $d^*$ is the
spatial dimension of the system.
Once an atom has been deposited it visits a region of linear size
$\ld$ and sticks inside that region after a time $\td\sim \ld^2/h$. 
The probability that another deposition occurs during this time in 
that region of size $\ld$ is 
\be
F\td\ld^{d^*} \sim (F/h)\ld^{2+d^*} \sim \ld^{2+d^*-\gamma}
\sim \ld^{-d^*} \, .
\label{eq_prob}
\ee

Since $\ld\gg 1$ such probability is very small, meaning that the sticking 
process of a 
new adatom is not influenced by atoms deposited later on.
These arguments lead us to consider an even simplified model,
which we are now going to describe.

Once an atom has been deposited, it immediately searches 
the surroundings for another atom or an island. If it exists, it sticks
to it, otherwise it sticks to the deposition site. Next, another
atom is deposited. The search will be implemented deterministically
through a capture area: the newly deposited atom (adatom in the following)
looks for the closest atom/island within a distance $\ell$. Each adatom
will therefore attach to an existing atom/island
or it will stick to the deposition site and collect atoms deposited later on 
in its capture area. 
It is to be noted that in the present model an adatom moves around 
only at the time of deposition, but never afterwards.
It is also important to stress that the size of the effective capture area 
reduces in the course of time, if newly deposited atoms
have their own capture area overlapping with it. The reason is that 
particles
at a distance $d$, with $\ell<d<2\ell$, do not capture each other, but their
capture areas overlap, and their capture distances 
along the joining line are reduced to $d/2$.

We are going to study the above model, both numerically and
analytically, in one spatial dimension ($d^*=1$). This choice is
due to the possibility to provide a full theoretical comprehension of 
numerical results, using two main analytic tools: capture equations 
and the analysis of the island-island distance distribution.

In the literature of submonolayer deposition, two models 
for representing islands are usually
found: `extended-islands' and `point-islands'.
In the former case, the physical size of an island increases proportionally
to the number of adatoms attached to it and its shape depends
on additional factors. In the latter case, islands have a physical size
equal to a lattice site whatever is the number of atoms they are made up.
The two models are statistically equivalent at small coverage 
(see next Section). We will use a `point-island' model
because of its simplicity.

A last remark relates to the physical interpretation of the capture 
length and to the novelty of our model.
If we refer to a system which displays a thermally activated
diffusion process, our capture length can be read as an
effective way to implement it. The idea 
is not new: it has been used, e.g., by Michael Biehl,
Wolfgang Kinzel and coworkers in several papers~\cite{Biehl} for studying
multilayer growth and it is related to the model of random deposition
with surface relaxation~\cite{BS}.
However, we are not aware of the application of the idea of
capture length to study the submonolayer regime.
On the other hand, the capture length may represent nonthermal
post deposition processes, as found, e.g., during the adsorption
of rare gas atoms on metal surfaces~\cite{rare_gas}.

This paper has three Sections, beyond the present Introduction.
In the next Section we provide a summary of the
main numerical results, while in Section~\ref{sec_quan} we derive
the analytical results and compare them to numerics.
In the final Section we discuss the hypotheses underlying our
model and suggest a possible interpolation between our model and
the `full diffusion' model.

Some remarks on notations are in order.
In the analytical calculations we generally use `dimensional' quantities, 
for example for the distance $d$ between islands. 
However, when we display numerical results and compare them with
analytics it may be useful to use reduce variables, for example
$x=d/\ell$.
In order to avoid any possible misunderstanding, we say that
$d,\bar d, \bar d_\infty, d_{nn},d_{nnn},d_l,d_r,y$ 
are all dimensional distances.
In particular, $\bar d$ is the average distance between neighbouring
islands and $\bar d_\infty$ is its asymptotic value for large
coverage (i.e., for large deposition time).
The only adimensional distances that are used throughout
the paper are $x$ and $\tilde d=\bar d_\infty/\ell$.

Finally, two probability distributions
for the distance will be introduced, $P_1$ and $P_2$. They refer
to the distance between first (nn) and second (nnn) nearest neighbouring
islands, respectively.

\section{Numerical Simulation}
\label{sec_qual}

Our model is defined as follows.
We choose randomly a lattice site and check if some atoms or islands
exist within a distance $\ell$. If they exist, the newly deposited
atom sticks to the closest atom/island, otherwise it sticks permanently to the 
deposition site. When an atom sticks to another atom, we get a new island. 

The model has two main features: it is a sequential model and random
deposition is the only source of noise. If $N_a$ atoms are deposited
on a substrate of $L$ sites, the ratio $\theta=N_a/L$ defines the coverage
$\theta$, which is also equal to the product of the deposition rate
and the deposition time, $\theta=Ft$. 
The statistical properties of the model depend on
the product between the capture area $c=2\ell+1$ and $\theta$:
this product is the average number of particles deposited in the area $c$.
All simulations have been done for $\ell\ge 10$, so that in the
following we can write $c\simeq 2\ell$ and introduce the
parameter $p=\theta\ell$.
Point-island and extended-island models have the same properties at small
coverage, let's say $\theta<0.2$. This condition can be satisfied
at any $p$, if the capture length is large enough, $\ell>5p$.

Let us now give a qualitative description of the growth morphology
followed by a summary of numerical results.
Since all quantities depend on $p=\theta\ell$ only, we can think
to keep $\ell$ fixed and vary $\theta$ or the other way round.
For pedagogical reasons we are describing the growth morphology with
increasing $\theta$.

At the very beginning, deposited atoms do not interact and each
atom or island grows in size according to the number of particles deposited
in its capture area $c=2\ell$. The average distance between atoms or islands
is very large. With increasing $\theta$, atoms or islands densities
increase and their capture areas start to overlap. From then on the growth
processes of different islands are no more independent. The average
distance is now between $\ell$ and $2\ell$. At large $\theta$, almost
all neighbouring islands have distances $d<2\ell$. The majority of
islands therefore increases in size according to the number of particles
deposited in a capture area equal to half the distance with
the left and right neighbours. Sometimes two neighbouring islands
have a distance $d>2\ell$ and the interval may be empty or contain
one atom. Densities of atoms and islands change because empty
intervals are filled by atoms and atoms are transformed into islands.
Let us now sum up the numerical results with the aid of the figures.

\begin{figure}[ht]
\includegraphics[width=12cm,clip]{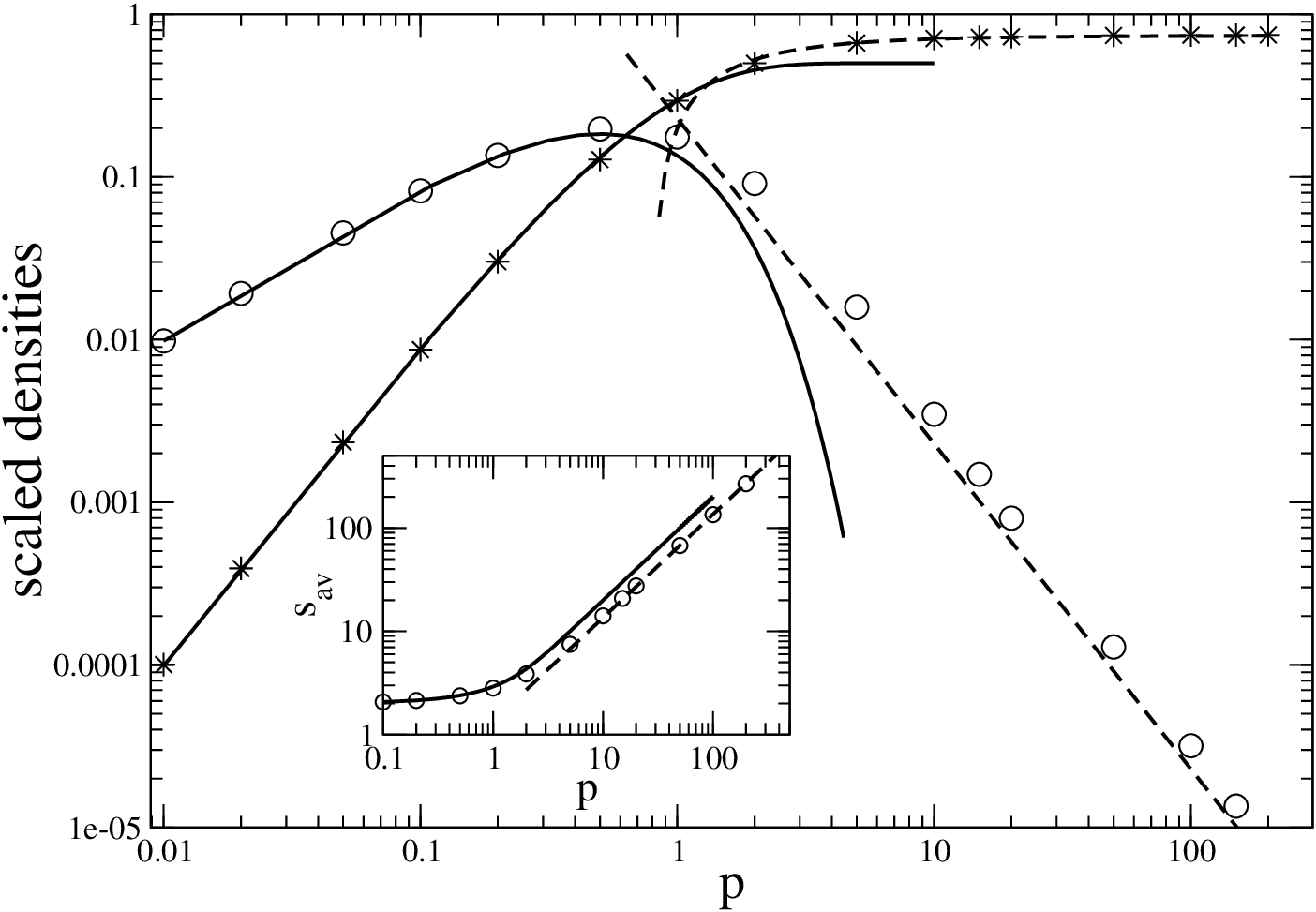}
\caption{
Numerical (symbols) and analytical (lines) results for
the adatom (circles) and island (stars) densities.
At small $p$ (full lines) adatom density varies as $\tilde n_1 =p\exp(-2p)$ 
and island density varies as $\tilde n_{is}= \frac{1}{2} -
(p+\frac{1}{2})\exp(-2p)$.
At large $p$ (dashed lines), $\tilde n_1$ decays as $1/p^2$ and
and $\tilde n_{is}$ approaches the asymptotic value $\tilde n_{is}^\infty$
as $1/p$ (see Eqs.(\ref{eq_n1_plarge},\ref{eq_nis_plarge})).
Inset: The average size of islands as a function of $p$.
Full line: small $p$ approximation. 
Dashed line: large $p$ approximation ($s_{av}=\tilde d p$).
All analytical results do not contain any fitting parameter.
}
\label{fig_densities}
\end{figure}

Fig.~\ref{fig_densities} summarizes all the main average quantities:
the densities of atoms (circles) and islands (stars) 
in the main figure and the average size of islands in the 
inset. At small $p$, the density of atoms ($\tilde n_1$)
increases linearly, of course, and the density of islands ($\tilde n_{is}$) 
increases quadratically (the tilde over a density means 
a reduced density, $\tilde n_{[]}=\ell n_{[]}$, see next Section). 
This behaviour is related to the specific features
of the nucleation process in the present model, which is due to the 
deposition of an atom in the capture area of an existing atom 
and it is not mediated by surface diffusion, as in epitaxial growth.
In the latter case $\tilde n_{is}$ increases with $\theta$ (i.e., with $p$)
as $\tilde n_{is}\approx \theta^\chi$, with an exponent~\cite{venables} 
$\chi=3$ in two
dimensions. Mean-field rate equations predict the same exponent in 
one dimension as well, but in this case the assumption $d \tilde n_{is}/
d\theta \approx \tilde n_1^2$ is wrong. The correct theory for
nucleation on top of a terrace~\cite{PC55} suggests 
$\chi=\frac{5}{2}$.

At large $p$, $\tilde n_{is}$ saturates but does not decrease
because coalescence of islands is not possible in a point-island model.
The limiting value,
$\tilde n_{is}^\infty$, is related to the reduced asymptotic average distance
between islands, $\tilde d=\bar d_\infty/\ell$, by the trivial relation
$\tilde n_{is}^\infty=1/\tilde d$. 
Island density converges
to $\tilde n_{is}^\infty$ as $1/p$, while adatom density decays
as $1/p^2$. These behaviours (see next Section) are due to the
existence of intervals slightly larger than $2\ell$, which are
`filled' by an adatom, which is subsequently transformed in a island.
The probabilities of the two processes differ: filling an interval 
$d\apprge 2\ell$ with
an adatom requires deposition on a small region
of size $d-2\ell$, while deposition on a region of size $\bar d_\infty$ 
is enough to make an island from an adatom. The process 
(adatom) $\to$ (island) is therefore quicker than the creation
of new adatoms:
this is the reason why $\tilde n_1$ decays more rapidly
than $\tilde n_{is}$ converges to $\tilde n_{is}^\infty$.
A quantitative, more rigorous analysis can be found in 
Section~\ref{sec_large}.

The inset of Fig.~\ref{fig_densities} gives the average size of islands,
$s_{av}$. It is of order two for small $p$. At large $p$, 
deposited atoms are shared among existing islands, and 
$s_{av}$ increases linearly with $\theta$, i.e. with $p$.

\begin{figure}[ht]
\includegraphics[clip,width=12cm]{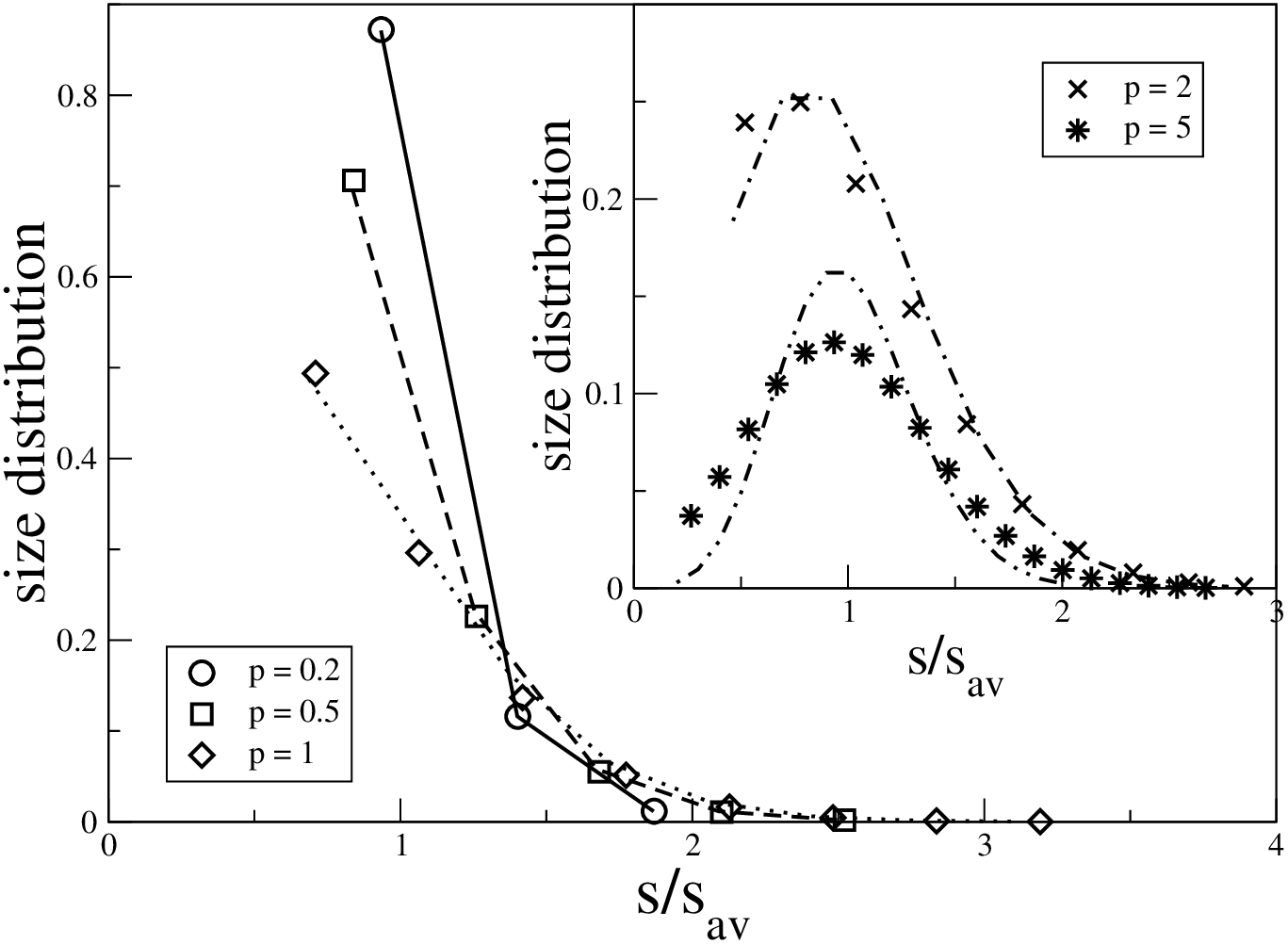}
\caption{The size distribution of islands for different values of $p$
(Main: $p=0.2,0.5,1$; Inset: $p=2,5$). 
Symbols: numerical results. Lines: analytical results according to the
small $p$ approximation, Eq.~(\ref{eq_ni}).
}
\label{fig_size_dist}
\end{figure}

\begin{figure}[ht]
\includegraphics[clip,width=12cm]{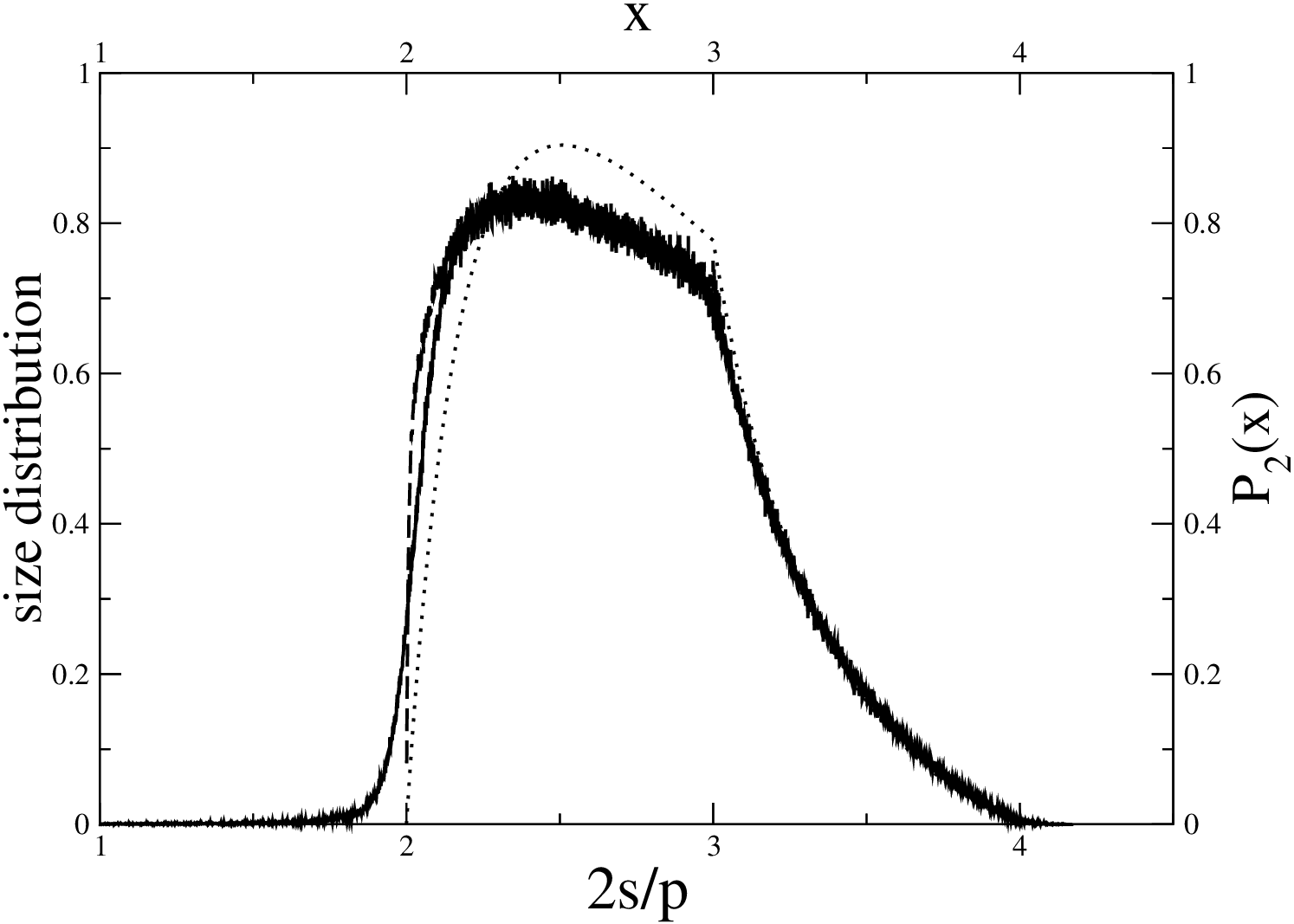}
\caption{Numerical results for the size distribution (full line),
as a function of $2s/p$.
Distance distribution between next-nearest-neighbouring islands:
numerical (dashed line) and analytical (dotted line) results
($x=d_{nnn}/\ell$).
}
\label{fig_P_s_nnn}
\end{figure}

Fig.~\ref{fig_size_dist} and Fig.~\ref{fig_P_s_nnn} plot the size 
distribution of islands, at small (Fig.~\ref{fig_size_dist})
and large (Fig.~\ref{fig_P_s_nnn}) $p$, respectively.
At small $p$, when different islands are almost 
independent, the size distribution is expected to follow a
Poisson distribution: this is confirmed by lines, which
reproduce well numerical data (symbols) till $p\apprle 2$.
At large $p$ an island grows in size because it collects atoms
deposited in a region of size $\frac{1}{2}(d_l + d_r)$,
where $d_{l,r}$ are the distances to the nearest left and right
neighbour. In this regime, the size distribution is therefore 
strictly related to the distance distribution between 
second nearest neighbouring (nnn) islands: this is proved in Fig.~\ref{fig_P_s_nnn}
by comparing the two (rescaled) distributions, plotted as full
line and dashed line, respectively. The dotted line is
the analytical nnn distance distribution ($P_2$), as derived 
from the island-island distance ditribution ($P_1$) under the assumption that 
there is no correlation between neighbouring distances 
(see the discussion in the next Section).

\begin{figure}[ht]
\includegraphics[clip,width=12cm]{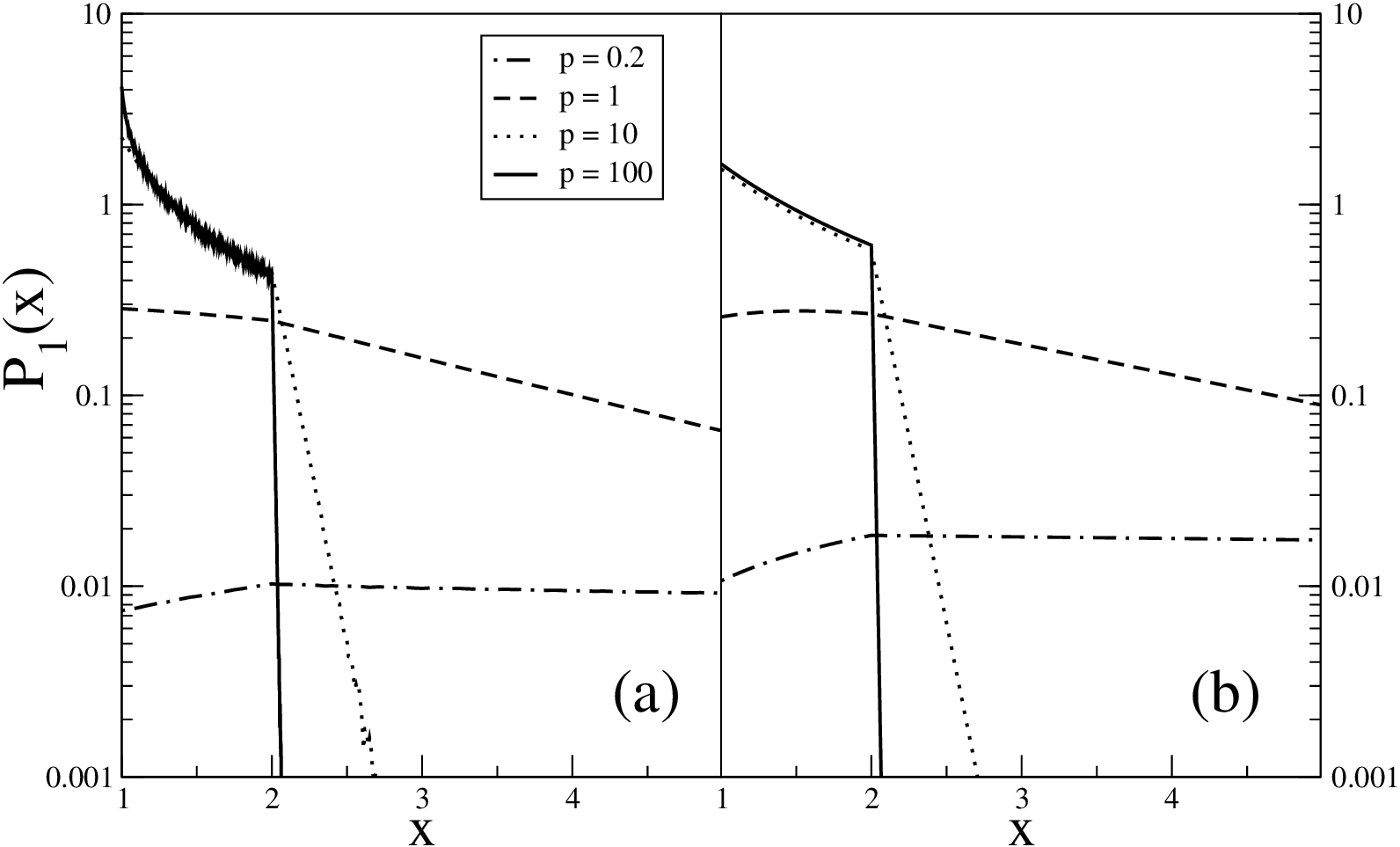}
\caption{
Simulation (a) and theory (b) for the island-island distance
distribution, as a function of $x=d/\ell$, for $p=0.2,1,10,100$.
}
\label{fig_distances}
\end{figure}

The previous discussion shows that the distance distribution
plays an important role to understand the statistical properties of the
model. In Fig.~\ref{fig_distances}a we report the 
distribution $P_1(x)$ of the normalized nn distance  $x=d/\ell$, 
for several values of $p$ (note the log-scale on the $y$ axis). 
Distances smaller than one (i.e., $d<\ell$)
are forbidden, of course. Two regimes are clearly visible, for 
$x$ smaller and larger than $2$. 
The reason is straightforward:
the capture areas of two neighbouring islands at distance
$x<2$ overlap, while they do not if $x>2$. This implies that
new islands can nucleate in between only in the latter case.

\begin{figure}[ht]
\includegraphics[clip,width=12cm]{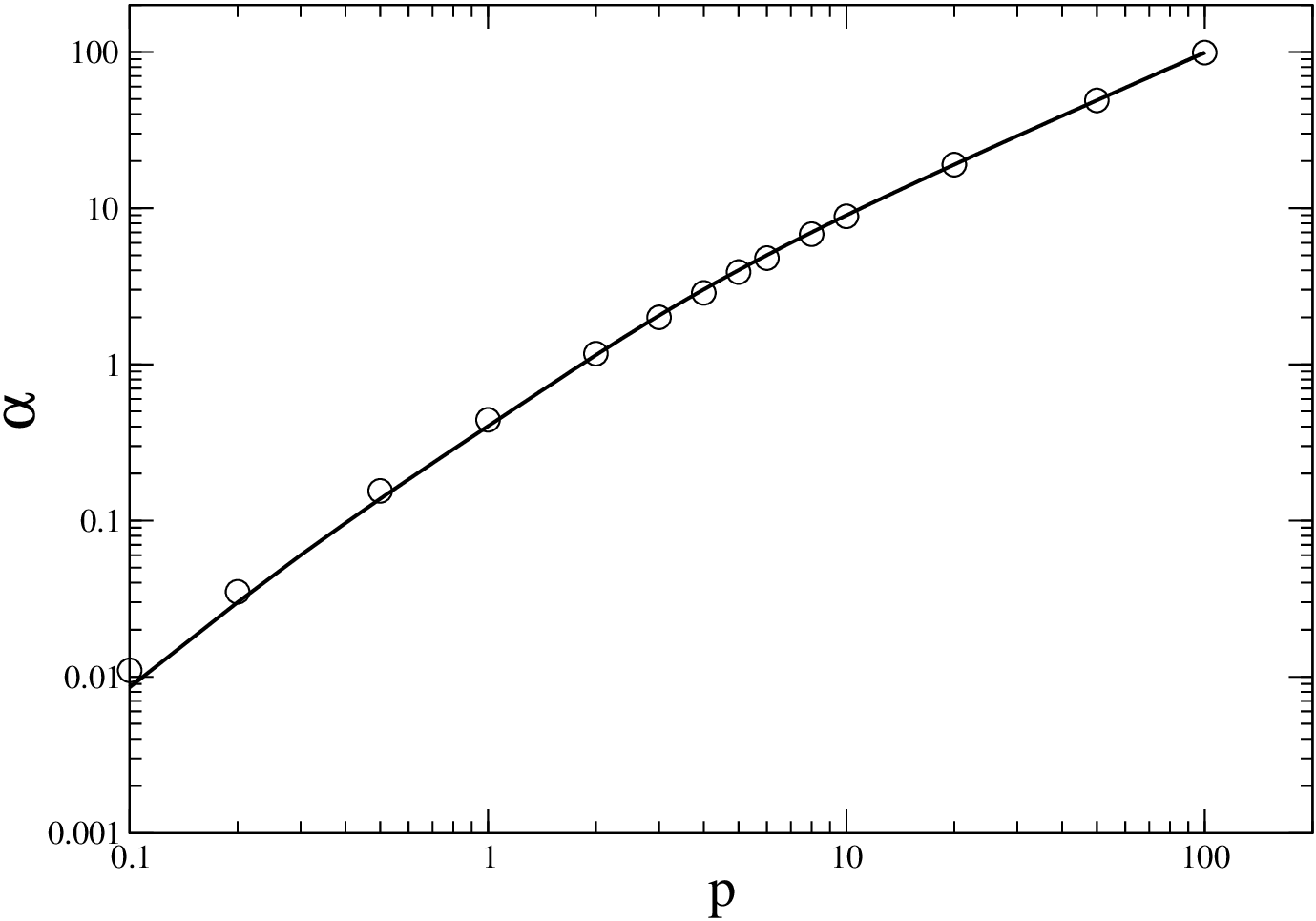}
\caption{Numerical (circles) and analytical (full line) results
for the prefactor $\alpha$, which characterizes the exponential
decreasing of the distance distribution $P_1(x)$, for
$x=d/\ell > 2$, $P_1(x) \sim \exp(-\alpha(p) x)$. $\alpha=p^2$ 
for small $p$ and $\alpha=p-1$ for large $p$.
}
\label{fig_alpha}
\end{figure}

\begin{figure}
\includegraphics[clip,width=12cm]{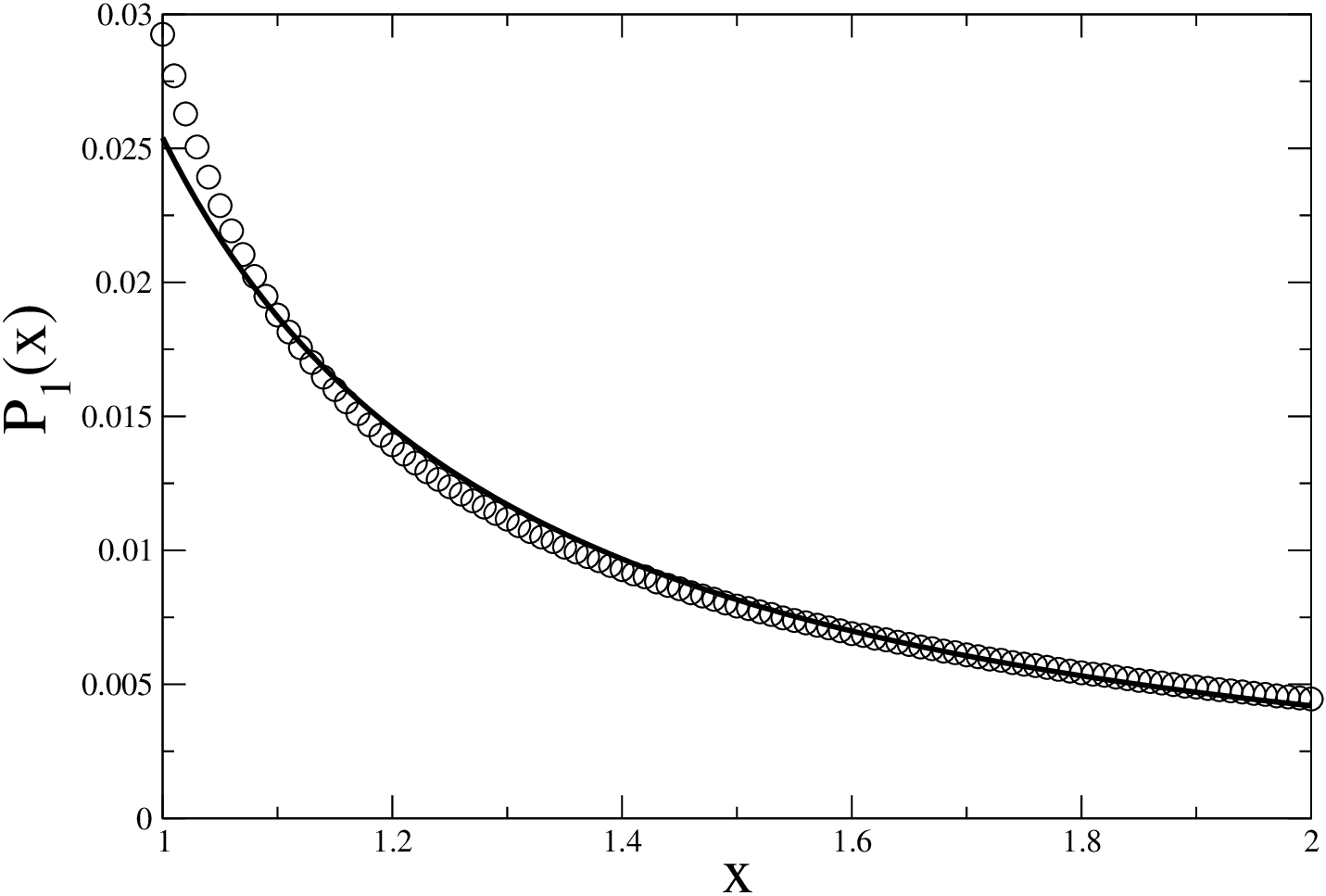}
\caption{Numerical (circles) and analytical (full line) results
for the `asymptotic' ($p=20$) distance distribution $P_1(x)$ in the interval
$1<x<2$, i.e. $\ell<d<2\ell$.
}
\label{fig_Pd1}
\end{figure}

For $x>2$, $P_1(x)$ decays exponentially, $P_1(x) \sim \exp(-\alpha x)$ with a
prefactor $\alpha$ depending on $p$ (see Fig.~\ref{fig_alpha}).
At smaller distances, $x<2$, $P_1(x)$ depends algebrically on $x$.
Furthermore, it is an increasing function at small $p$ and a
decreasing one at large $p$. For very large $p$, it converges to 
a limiting shape (see Fig.~\ref{fig_Pd1}), except for 
$x$ very close to one.

\section{Quantitative Analysis}
\label{sec_quan}

\subsection{Densities and size distributions for small $p$}
\label{sec_small}

In the following $n_1$ is the density of adatoms (adatoms per lattice site),
$n_s$ ($s\ge 2$) is the density of islands of size $s$, and
$n_{is}=\sum_{s\ge 2} n_s$ is the total density of islands.
The quantities $\tilde n_{[]} =\ell n_{[]}$ are the `reduced' densities
and they mean the number of adatoms/islands per capture length.
In the limit of small $p$ deposited particles do not interact because
their distance is larger than $2\ell$ and each adatom or island has
a capture area equal to $2\ell$. In this limit it is possible 
to write down the following capture equations ($p=\theta\ell$):
\bea
{d \tilde n_1 \over dp} &=& 
1 -2(\tilde n_{is} + 2\tilde n_1) 
\label{re_n1}  \\
{d \tilde n_{is} \over dp} &=& 2\tilde n_1
\label{re_nis}
\eea
which can be easily solved, giving
\be
\tilde n_1 = p\exp(-2p) \, ,~~~~
\tilde n_{is} = \hbox{${1\over 2}$} - (p+ \hbox{${1\over 2}$})\exp(-2p) \, .
\label{eq_n1_nis}
\ee

These expressions, that are consistent for any $p$, 
are reported in Fig.~\ref{fig_densities} as
full lines and compared with numerical results (symbols). 
Comparison is almost perfect till $p\approx 1$.
For larger values overlapping of capture areas is relevant and
the above approximation is no more valid.
The main defect of (\ref{eq_n1_nis}), due to the assumption
of a constant capture area, is that $\tilde n_1$ and 
$\frac{1}{2}-\tilde n_{is}=\tilde n_{is}^\infty-\tilde n_{is}$
decrease exponentially for $p\to\infty$, instead of that in a power law.
The limit of large $p$ will be treated in Section~\ref{sec_large}.

The equations for the density evolutions of islands of any size
$s\ge 2$ are
\be
{d \tilde n_s \over dp} = 2(\tilde n_{s-1} -\tilde n_s) 
\label{re_ni}
\ee
and they can be solved recursively,
\be
\tilde n_s = {1\over 2} {(2p)^s\over s ! } \exp(-2p) ~~~ s\ge 1
\label{eq_ni}
\ee
showing that at small coverage the size distribution is Poisson-like.
In Fig.~\ref{fig_size_dist} we compare the above analytical expressions
with numerical results: they match very well till $p\approx 1$.
The average size $s_{av}$ of islands ($s \ge 2$) is calculated as
\be
s_{av} = \frac{\sum_{s=2}^{\infty} s \tilde n_s}{\tilde n_{is}}
= 2p {\exp(2p) - 1 \over \exp(2p) -(1+2p) } \, .
\label{eq_sav}
\ee

This small $p$ approximation for $s_{av}$ is reported in the inset of
Fig.~\ref{fig_densities} as a full line. It also reproduces the correct
linear behaviour at large $p$ ($s_{av}=ap$), but with a wrong factor $a$.
The reason is that the small $p$ approximation gives an
asymptotic value for the island density, $\tilde n_{is}^\infty=
\frac{1}{2}$, which is smaller than the actual value
(compare stars and full line in Fig.~\ref{fig_densities});
therefore $a=1/\tilde n_{is}^\infty$ is overestimated.

Finally, expression (\ref{eq_ni}) for $\tilde n_s$ 
might be slightly improved
for $p\apprge 1$ by replacing $(2p)$ with a parameter $\lambda$
in Eqs.~(\ref{eq_ni}) and (\ref{eq_sav}). This parameter
should be determined by Eq.~(\ref{eq_sav}) using the actual numerical
value for $s_{av}$.

\subsection{Island-Island distance distribution}
\label{sec_distances}

In Fig.~\ref{fig_distances}a we plot the distribution of distances
between nearest neighboring
islands as a function of the `reduced' distance $x=d/\ell $.
The existence of two regimes separated at $x=2$
is due to the possibility of creating 
a new island within an interval, only if $d>2\ell$.
The weight of the two regimes changes with $p$, because the fraction of 
distances $d<2\ell$ increases from zero to one as $p$ increases.

For $x>2$ the distribution has an exponential behaviour,
$P_1(x)\sim\exp(-\alpha x)$. In this regime the capture areas of 
two neighbouring islands do not overlap and $P_1(x)$ is just
the probability that no island nucleation occurs in time
$\theta$ in the interval of size $d$ between the two islands\cite{note1}. 
If $\omega(t)$ is the nucleation probability per unit time and
unit length, at time $t$,
neglecting correlations of nucleation events we obtain
\be
P_1(x) = \exp\left[ -d\hskip -.1cm 
\int_0^\theta \hskip -.1cm dt \, \omega(t)\right] \, .
\label{eq_Pd2}
\ee

Nucleation occurs when an atom is deposited in the capture area of an 
already existing atom.
If a deposition event increases the (total) capture area by $\Delta$,
then
$d\omega/dt = (1-\omega)\Delta$, i.e. $\omega(t)=1-\exp(-\Delta t)$.
Inserting the last expression in (\ref{eq_Pd2}) 
and introducing $\delta=\Delta/\ell$, we obtain
\bea
\alpha(p) &=& \int_0^p d\tau (1-e^{-\delta \tau})  \\
&=& p - (1-e^{-\delta p})/\delta \, .
\label{eq_alpha}
\eea

In fact, the quantity $\delta$ is not constant, 
but it decreases in time, i.e. with $p$, being equal
to two for $p\to 0$. In Fig.~\ref{fig_alpha} we compare numerical
data for $\alpha$ with Eq.~(\ref{eq_alpha}), assuming
$\delta=1+\exp(-p)$, which gives the correct limits $\delta\approx 2$ for 
small $p$ and $\delta\approx 1$ for large $p$.\cite{note2}
The agreement is fairly good.

If $x<2$ the determination of $P_1(x)$ is more
difficult because of the `interaction' between the capture zones of the two
neighbouring islands. The existence of two islands in $y=0$ and
$y=d$ after a time $\theta$ can be depicted as follows. 
If we assume that the (first) atom in zero is deposited at time $t_1$
and the atom in $d$ is deposited at time $t_2>t_1$, we require
that no deposition takes place in a region $(d+2\ell)$ during time
$t_1$, no deposition in a region $d$ between $t_1$ and $t_2$, at least
one deposition in $2\ell$ between $t_1$ and $\theta$, at least one
deposition in $d$ between $t_2$ and $\theta$\cite{note3}.
Integrating over $t_1$ and $t_2$ the product of all the above
probabilities provides $P_1(d)$ in the hypothesis that neighbouring
intervals have no influence, 
\bea
P_1(d)
 &=&
\int_0^\theta dt_1 \exp(-2\ell t_1) \int_{t_1}^\theta
d t_2
\exp(-d t_2) \times 
\nonumber\\
&& \times [1- \exp (-d(\theta-t_2))] 
[1- \exp(-2\ell(\theta-t_1))]
\nonumber\\
&=& \exp(-(d+2\ell)\theta) [I_1/d - I_2] \label{eq_Pd1}
\eea

where
\bea
I_1 &=& {\exp((2\ell+d)\theta)-1\over 2\ell+d} -
{\exp(2\ell\theta)-1\over 2\ell} -
{\exp(d\theta)-1\over d} + \theta \\
I_2 &=& \left(
2\ell\theta\exp(2\ell\theta) -\exp(2\ell\theta) - 2\ell^2\theta^2 + 1\right)
/4\ell^2 \, .
\eea

Prefactors are missing in Eqs.~(\ref{eq_Pd2},\ref{eq_Pd1}). They 
are determined imposing the continuity of $P_1(d)$ in $d=2\ell$ and
its normalization in $(\ell,\infty)$.
The result of this procedure is shown in Fig.~\ref{fig_distances}b,
whose qualitative comparison with simulation results 
(Fig.~\ref{fig_distances}a) is rather satisfying.
The quantitative comparison is good for small $p$, but is not
for large $p$. The reason is simple: 
for large $p$, the average distance $\bar d$ between islands
is smaller than $2\ell$, which means that neglecting the left and
right neighbours of the two islands located in $y=0$ and $y=d$ 
is no more correct.
It is possible to take them into account in a `mean field
approximation', by assuming the existence of such neighbours
at a fixed distance $\bar d$. At small $p$, $\bar d>2\ell$ and this
refinement is inessential, but for large $p$ is relevant.

The calculation proceeds along the same lines leading to (\ref{eq_Pd1})
with the difference that now capture areas are limited by the
presence of two additional islands in $y=-\bar d$ and $y=d+\bar d$.
If we limit ourselves to the case $p\gg 1$, the refined procedure gives
\be
P_1(d) = {P_0\over (d+\bar d_\infty - 2\ell)(d+2\bar d_\infty -2\ell)} \, ,
\label{eq_p1d} 
\ee
where $P_0$ and $\bar d_\infty$ should be determined
self-consistently through the conditions
\be
\int_\ell^{2\ell} dy P_1(y) =1 ~~~,~~~
\int_\ell^{2\ell} dy yP_1(y) = \bar d_\infty \, .
\ee

In Fig.~\ref{fig_Pd1} we compare the analytical and numerical results
for $P_1(d)$ in the regime $d<2\ell$, using the reduced distance $x=d/\ell$. 
A couple of remarks are in order.
First, the analytical curve has no fitting parameter.
Second, the inverse of the reduced average distance $\tilde d=
\bar d_\infty/\ell$ has the value $1/\tilde d =0.738$, 
determined self-consistently from the above procedure.
This value agrees very well
with the asymptotic island density for large $p$,
$\tilde n_{is}^\infty \approx 0.75$.
The value of $\tilde d$ also allows to draw the dashed line in the
inset of Fig.~\ref{fig_densities}, which gives the theoretical prediction
$s_{av}=\tilde d p$ for the average size of islands at large $p$.

A final comment concerns the shape of $P_1(x)$ for $x<2$, in the limit
$p\to\infty$. The curve with circles in Fig.~\ref{fig_Pd1} refers
to $p=20$. Numerical results show that a limiting shape does exist
for $x\apprge 1.1$, while $P_1(1)$ seems to have a logarithmic divergence.
This behaviour is due to distance-distance correlations which are
not taken into account by the mean-field approximation
leading to Eq.~(\ref{eq_p1d}).

\subsection{Size distributions and adatom/island densities for large $p$}
\label{sec_large}

Fig.~\ref{fig_P_s_nnn} shows that the asymptotic size distribution $\tilde n_s$
(full line) agrees well with the distance distribution $P_2(x)$
(dashed line) between next-nearest-neighbouring islands ($x=d_{nnn}/\ell$). 
The reason of that agreement is easily explained, because the
density of islands is almost constant for $p\gg 1$.
Nearly all deposited atoms are captured by preexisting islands,
which grow according to their capture area. The capture area of each
island is just half the distance with its left neighbour plus half the
distance with its right neighbour, $\frac{1}{2}(d_l+d_r)=\frac{1}{2}d_{nnn}$.
Therefore, for large $p$ the size $s$ grows accordingly to
the relation $s=d_{nnn}\theta/2$ and the distance distribution between 
nnn islands is equivalent to the size distribution of the islands.

In order to compare $\tilde n_s$ and $P_2(x)$, 
the size distribution is plotted as a function of $2s/p$. 
The two curves slightly differ  for $x\apprle 2 (s\apprle p)$:
$P_2(x)$ vanishes for $x<2$, while $\tilde n_s$ has a tail at small
size. This tail disappears in the limit $p\to\infty$, but very
weakly, because $\tilde n_{is}$ converges to $\tilde n_{is}^\infty$
as $1/p$ only (see Fig.~\ref{fig_densities}).

It is interesting to compare $P_2(x)$ as derived from simulations with the
nnn distance distribution, as derived from
the nn distance distribution, $P_1(x)$, in the hypothesis that
neighbouring intervals are independent (Fig.~\ref{fig_P_s_nnn}, dotted
line).
From the general relation
\be
P_2(d) = \int dy P_1(y) P_1(d-y) \, ,
\label{eq_P2d}
\ee
for large $p$ we have
\be
P_2(d) = P_0^2 \int {dy\over
(y+\bar d_\infty -2\ell)(y+2\bar d_\infty -2\ell)(d-y +\bar d_\infty -2\ell)
(d-y +2\bar d_\infty -2\ell) }
\ee
where $P_0$ is the (known) normalization factor for $P_1(y)$.
The qualitative behaviour of $P_2(d)$ can be understood
as follows. For large $p$, $P_1(y)$ is practically non-vanishing only in
the region $\ell \le y \le 2\ell$, so that $P_2(d)$ does not vanish
for $2\ell \le d \le 4\ell$. The shape of $P_2(d)$ is determined
by two factors: $P_1(y)$ is a continuously decreasing
function, and different $d$ have a different `weight' $\rho(d)$. 
If we rewrite (\ref{eq_P2d}) as a two-dimensional integral, 
$P_2(d) = \int_\ell^{2\ell} dy_1 dy_2 P_1(y_1) P_1(y_2)\delta(d-y_1-y_2)$,
the weight is just the quantity 
$\rho(d) = \int_\ell^{2\ell} dy_1 dy_2 \delta(d-y_1-y_2)$,
which has a symmetric maximum in $d=3\ell$ and vanishes at the
extremities $d=2\ell,4\ell$. The non-analytic maximum is the responsible
for the change of slope of $P_2(d)$ for $d/\ell=3$, while $\rho(2\ell)=
\rho(4\ell)=0$ explain the vanishing of $P_2(d)$ at the same points.

As $d$ increases from two to four $\ell$, the functions $P_1$
in (\ref{eq_P2d}) are evaluated, in average, at increasing values
of the argument. Therefore, for $d>3\ell$ both the weight $\rho(d)$ and
the product of $P_1$ are decreasing functions of $d$: this explains
the fast decreasing of $P_2(d)$ in that interval.
For $d<3\ell$ the weight is an increasing function vanishing in
$2\ell$ and the product of $P_1$ is a decreasing function:
this justifies the presence of a maximum of $P_2(d)$ in that interval.

Comparison of dashed and dotted lines in Fig.~\ref{fig_P_s_nnn} shows
a major difference in the region $d_{nnn}\apprge 2\ell$,
i.e. for $d_{nn}\apprge\ell$. This disagreement is due to two reasons.
First, the theoretical $P_1(d)$ underestimates the true island-island
distance distribution for $d\simeq\ell$ (see Fig.~\ref{fig_Pd1}).
Second, Eq.~(\ref{eq_P2d}) assumes there are no correlations between
neighbouring intervals, which is not the case.

Let us finally discuss the adatom and island densities in the limit
of large $p$ (Fig.~\ref{fig_densities}, dashed lines).
In the large $p$ regime, the surface is a sequence of islands separated
by distances $d<2\ell$ (with average value $\bar d_\infty$).
Rarely are there intervals with $d>2\ell$: some of them, equal in
number to $I_0$, are void, the others ($I_1$) contain one atom.
The total number of intervals is approximately 
equal to $I_t=L/\bar d_\infty$ and it
is assumed to be constant, because $I_t\gg I_0,I_1$.
The number of atoms is equal to $I_1$.

If $N_a$ is the number of deposited atoms ($N_a=\theta L$), $I_0$ and
$I_1$ satisfy the following equations
\bea
{dI_0\over dN_a} &=& - {\Delta y\over L}I_0 \label{eq_I0} \\
{dI_1\over dN_a} &=&  {\Delta y\over L}I_0 - {\bar d_\infty
\over L}I_1 \, , 
\label{eq_I1}
\eea
where $\Delta y$ is the `active' region of an interval $d$ larger
than $2\ell$ (active means that a deposition event in such region
creates a new adatom). We can evaluate it as $\Delta y=\langle d\rangle_> 
-2\ell$, where the average $\langle d\rangle_>$ is performed 
on intervals $d>2\ell$ only.

For $d>2\ell$ (and large $p$) the distance distribution between islands is
\be
P_1(d) = P_1(2\ell) \exp[-\theta(d-2\ell)] \, ,
\ee
with $P_1(2\ell)$ which can be determined by Eq.~(\ref{eq_p1d}).

The integration of the previous equation gives the probability
that a distance $d$ is larger than $2\ell$,
\be
P(d>2\ell) = {c_0\over p} \, ,
\ee
with $c_0=[2\tilde d\ln({2\tilde d-1\over 2\tilde d -2})]^{-1}$.
Much in the same way we can determine $\Delta y$,
\be
2\ell+\Delta y = { \int_{2\ell}^\infty dy \; y e^{-\theta y} \over
\int_{2\ell}^\infty dy  e^{-\theta y} } \; ,
\ee
which gives $\Delta y=1/\theta$.

Eqs.~(\ref{eq_I0},\ref{eq_I1}) can now be rewritten as
\bea
{dI_0\over dp} &=& - {I_0\over p} \\
{dI_1\over dp} &=& {I_0\over p} - \tilde d I_1
\eea
whose solutions are 
\bea
I_0(p) &=& {c_1\over p} \label{sol_I0}\\
I_1(p) &=& c_1\tilde d\left[ {1\over (\tilde d p)^2} +
\sum_{n=3}^\infty { (n-1)! \over (\tilde dp)^n} \right] \approx
{c_1\over\tilde d}{1\over p^2} \, .
\label{sol_I1}
\eea

Eq.~(\ref{sol_I1}) proves that the adatom density vanishes
as $1/p^2$. Since $I_0\sim 1/p$, island density and the total
density (adatoms$+$islands) both converge to 
the asymptotic value with corrections of order $1/p$.
Finally, we can determine analytically $c_1$,
because 
\be
{c_0\over p}=
P(d>2\ell) = {I_0+I_1\over I_t} = {c_1\bar d_\infty\over L}{1\over p}
\ee
so that $c_1=(L/\bar d_\infty)c_0$ and
\be
\tilde n_1 = \ell {I_1\over L} = {1\over 2\tilde d^3\ln(
{2\tilde d-1\over 2\tilde d-2} ) } \,
{1\over p^2} \, .
\label{eq_n1_plarge}
\ee

In Fig.~\ref{fig_densities} we compare numerical data (circles)
with the previous expression for $\tilde n_1$ (decreasing dashed line): 
the slope (-2) is pretty correct, but
the analytical prefactor is of order 0.23, to be compared with the numerical
value, 0.34.
Along the same lines, it is possible to determine $\tilde n_{is}$ for large 
$p$:
\be
\tilde n_{is} = {1\over\tilde d} -
{1\over 2\tilde d^2 \ln\left( {2\tilde d -1\over 2\tilde d -2}\right) }
{1\over p}\left(1 + {1\over\tilde dp} \right ) \; .
\label{eq_nis_plarge}
\ee

This analytical expression is compared succesfully to numerical data
in Fig.~\ref{fig_densities} (see the dashed line superposing to
stars).

\section{Comments}

The results for our model can be compared to the `full diffusion' model,
where nucleation and aggregation are due to the thermally
activated diffusion process. Even if a detailed comparison
is postponed to a future paper~\cite{extended} which will 
extend our calculations and simulations to two dimensions
and to a sequential diffusion model (see below),
some comments are in order here.

In Section~\ref{sec_qual} we already stressed an important difference
concerning the small coverage $\theta$
 behavior of island density, $\tilde n_{is}\approx
\theta^\chi$, with an exponent $\chi$ which is different for the
two models (in $d^*=2$ the difference is even more relevant,
because $\chi=2$ for our `no diffusion' model, while
$\chi=3$ for the `full diffusion' model). Another difference
concerns the shape of the size distribution of islands.
The origin of these differences should be traced back to
Eq.~(\ref{eq_prob}), which gives a rough evaluation of the
probability that a third atom intervenes during the nucleation
or aggregation process of a given atom. As a matter of fact, this
criterion has two weak points, both related to the actual
meaning of the diffusion length.

First, in the early growing regime atoms
may travel a distance larger than $\ld$ before being
incorporated: $\ld$ can be correctly defined as a typical distance
only in the regime where the island distance is approximately constant.
Second, two atoms may stick together
even if their `initial' distance is larger then $\ld$, or, similarly,
they may not stick even if their distance is smaller than $\ld$:
$\ld$ is an average quantity and taking it as a capture length 
kills diffusion noise.

In order to evaluate the effect of the two features of our model,
sequentiality and absence of diffusion noise, we plan to study 
an intermediate model, which we can call `sequential diffusion' model.
This model is still sequential, but nucleation/aggregation
does not occur deterministically. The capture length recipe
is replaced by allowing a deposited atom to walk a fixed 
maximum number of random hops. Preliminary simulations~\cite{extended} 
in one dimension show that this model has statistical properties 
more similar to the `full diffusion' model.

\section*{Acknowledgements}
Authors acknowledge Thomas Michely to have pointed out 
Ref.~7).
PP would like to thank the Japan Society for the Promotion of Science 
for a research grant which supported his stay at Keio University
in Spring 2004.

%


\begin{thebibliography}{99}
\bibitem{note1}
As a matter of fact we should evaluate the probability of no nucleation
in a interval of size $d-2\ell$. Because of the exponential behaviour
we can replace $d-2\ell$ with $d$.
\bibitem{note2}
The limit $\delta=2$ (i.e., $\Delta=2\ell$) for $p\to 0$ is
straightforward.  The limit $\delta=1$ for large $p$ is necessary
to obtain $\alpha=p-1$ in that limit.
\bibitem{note3}
This is not quite exact. We should require to have at least one
deposition in the capture area of the first atom ($y=0$) between $t_1$ and
$\theta$ and one deposition in the capture area of the second atom ($y=d$) 
between $t_2$ and $\theta$. The point is that the capture area of
the first atom changes at time $t_2$. Our approximate expression,
which is assumed for the sake of simplicity, uses the correct
capture area for $t<t_2$ and the correct total capture area
for $t>t_2$.
\end{thebibliography}
\end{document}